\documentclass{aa}
\usepackage{amssymb}
\usepackage{graphicx}

\newcommand{\etal}{et al. }

\begin{document}
 
\title{ISOCAM Mid-infrared spectroscopy and NIR photometry of the HII complex N4 in LMC.
\thanks{Based on observations with ISO, an ESA
project with instruments funded
by ESA member states (especially the PI countries: France, Germany, the
Netherlands and the United Kingdom) and with the participation of ISAS and
NASA.}}
\author{A. Contursi\inst{1}\and
M Rubio\inst{2}\and
M. Sauvage\inst{3}\and
D. Cesarsky\inst{1,5} 
R.Barba\inst{5}\and
F. Boulanger\inst{6}
}
\offprints{Contursi A.,
\email {contursi@mpe.mpg.de}}

\institute{Max-Planck-Institut f\"{u}r Extraterrestrische Physik (MPE), 
Postfach 1312, 85741 Garching, Germany \and	
Departamento de Astronom\'ia Universidad de Chile, Casilla
36--D, Snatiago, Chile\and
CEA/DSM/DAPNIA/Service d'Astrophysique, CE Saclay, 91191 Gif-sur-Yvette Cedex, France,
France\and
Departamento de F\'isica, Universidad de La Serena, Chile\and
Institut d'Astrophysique Spatiale, Bat. 121, Universit\'e Paris
XI, 91450 Orsay CEDEX, France}

\date{Received 22 August 2006/ Accepted }
 
\abstract{  We present the analysis of ISOCAM-CVF and J,H and K$_s$ 
photometry data of the HII region complex N4 in the Large Magellanic Cloud (LMC).}
{ The aim is twofold: 1) to study the connection between the interstellar medium
and the star content of this region;
2) to investigate the effects  of the lower than galactic metallicity  on dust
properties.}
{A dust features -- gas lines -- continuum fitting technique on the whole ISOCAM-CVF data cube,
allows the production of images in each single emission and the detailed analysis
of dust (both continuum and bands), and ionized gas. The Near Infrared (NIR)  photometry provides, for the first time,
 information on the stellar content of N4. }
{ The Mid--Infrared (MIR) spectral characteristics of N4 are those expected for
an HII region complex, i.e. very similar to those  observed in galactic HII regions. 
The images in  single dust feature bands and gas lines clearly  show  
 that the HII region core is completely devoid of
the carriers responsible for the Aromatic Features (AFs).  On the other hand, the ionized gas arises almost  
completely in this dust cavity,
 where also the two main exciting stars of N4 are located. HII region models from Stasi\'nska
(1982) predict an   HII region size which  
 corresponds to  the observed size of the dust  cavity. \\
 We find evidences that the effect of  lower than Galactic  metallicity (although not extreme as in the case of LMC) 
 on the carriers responsible for the AFs, is not to prevent their 
 formation or to modify  their chemical properties,  but to enhance their destruction by 
 the high and  hard  interstellar radiation field. We argue that this
 is the dominant process   responsible for the absence of AFs   in the HII
 region core. We show that this mechanism is    more efficient on smaller dust particles/molecules 
 thus affecting the dust-size distribution. We argue that effects on dust--size distribution, 
 rather than the  different dust properties due to a lower metallicity, 
   should be taken into account when analyzing 
 more distant relatively low metallicity galaxies.\\
  Finally, the analysis of the 
 stellar content of N4 reveals 7 stars: 4   reddened O MS stars  and  three stars with envelopes. In particular, one of these,  
 seems to be  an Ultra Compact HII region containing an embedded Young Stellar
 Object.}{}


\maketitle
%\markboth{A. Contursi \etal: ISOCAM Mid-infrared spectroscopy and NIR photometry of N4 in LMC}{The}

\section{Introduction}
N4 (Henize \cite{Henize}) is an HII  complex in the north-west part of the Large Magellanic Cloud (LMC).
 The physical characteristics of the ionized gas and associated molecular cloud,  
 have been extensively studied
 by  Heydari-Malayeri and Lecavelier des Etangs (\cite{Heydari}) 
 through optical and submillimeter line spectroscopy. 
 
\begin{figure}
\includegraphics[width=0.5\textwidth,height=0.5\textwidth]{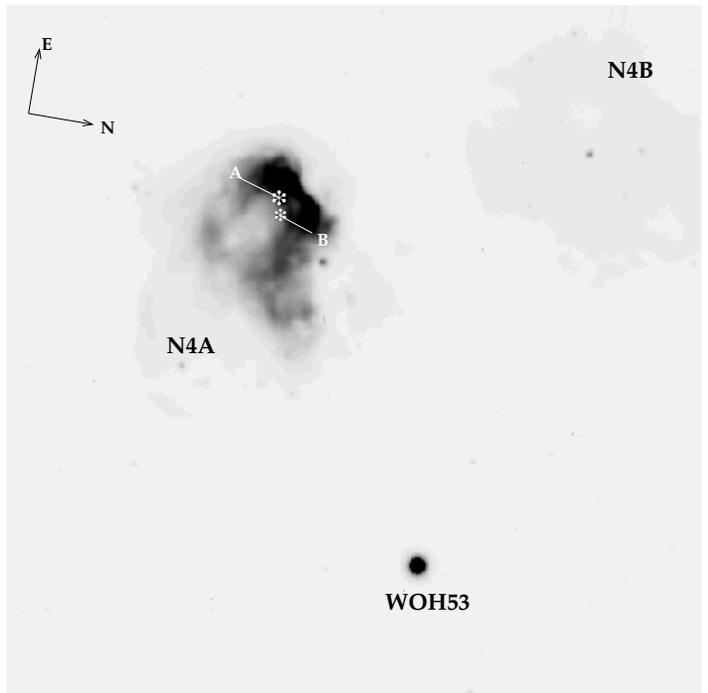}
\caption{H$\alpha$ emission of the  LMC HII complex N4A and N4B (from Heydari-Malayeri
and Lecavelier des Etangs 1994). The image has been rotated to the same orientation as the
ISOCAM--CVF images.}
\label{Figure1}
\end{figure}

The H$\alpha$ emission of N4 (Figure \ref{Figure1}, kindly provided by 
Heydari-Malayeri) clearly shows that this  HII complex is 
composed of two ionized nebulae: N4A  and N4B.
In this paper we will concentrate only on N4A.
This is  brighter and younger than N4B. It is
 composed of a bright dense front to its north-east part  
 and a more diffuse low surface brightness, partially density bounded, component
 to its south part.
  The main ionizing sources
 in N4A are two stars not visible at  optical wavelengths (marked in Figure
 \ref{Figure1} as A and B).  
 Heydari-Malayeri and Lecavelier des Etangs (\cite{Heydari})  calculated
 that the main ionizing sources correspond to  either one 60M$\odot$ star 
 or two 40 M$\odot$ stars. \\
 The average extinction in N4 derived from optical gas emission is quite
 low, A$_V$=0.4 although it is presumably higher in the north-east front where the gas density reaches 
 its maximum.
  The parental molecular cloud is formed by two spatial components: one peaking on N4A and the other
  peaking to the east  of the HII region. The molecular
  cloud has two velocity components: a high velocity optically thick red component and a thin blue molecular
  sheet. This sheet is   probably pushed
   towards us  by the HII region pressure (Heydari-Malayeri and Lecavelier des Etangs \cite{Heydari}).
  
A complementary study on the hot dust emission in  N4A has been done by  Contursi \etal (\cite{Contursi}) using
broad band ISOCAM filters centered at 7 and 15 $\mu$m. At these wavelengths, the emission is dominated by the  
Aromatic Features seen in Emission (AFEs)  between 3 and 10 $\mu$m 
and a continuum steeply rising at wavelengths longer than  $\sim$10 $\mu$m. 
The overall Mid--Infrared (MIR)  morphology in N4A
  is generally quite similar to the CO emission, although the MIR and CO peaks do not exactly coincide. 
  The  strongest MIR emission comes from the dense north-east front as expected, since here the density of the matter is the
  highest. The 15/7 $\mu$m ratio peaks in the region of N4A where the two main exciting stars are supposed
  to be. Its high value, typical for an HII region,   indicates that the emission at 15 $\mu$m  is dominated by a strong
  continuum and that the emission at $\sim$ 7 $\mu$m is significantly reduced. Contursi \etal (\cite{Contursi}) suggested
  that this decrease is primarily due to  destruction of the AFEs carriers 
  in hard and strong Interstellar Radiation Fields (ISRF). 
  
  This analysis was part of a wide ISOCAM campaign aimed at studying the interstellar medium
  (ISM) properties in the Magellanic Clouds with particular emphasis on the role
  played by metallicity, taking full   advantage of the unprecedent (at that time) spatial resolution   provided by ISO at these wavelengths
  and to the proximity of the Magellanic Clouds.  
  One of the outcomes of the broad band ISOCAM imaging study of N4
   was that this region appears as a classical  Galactic HII complex, composed  mainly  of three phases: 1)
  The HII region itself close to the exciting stars corresponding to the maximum of the 15/7$\mu$m ratio; 2) the interface between 
  the HII region and the molecular cloud (Photon--Dominated Region or PDR) where the MIR emission reaches its maximum due to both
  high dust density and relatively high ISRF; 3) the parental molecular cloud, where the MIR
  emission is still dominated by AFEs probably arising from the outskirts of the molecular clouds and excited mainly by the
  general LMC ISRF.
  
 In this paper we present a more detailed study of the MIR emission 
 properties of N4A analyzing  a very high signal to noise
 spectro--imaging data set carried out with the Circular Variable Filter (CVF) 
 on ISOCAM. Thanks to the spectral capabilities, 
  we are able to directly show that our previous physical interpretation 
  of the broad band MIR photometry data of N4, 
  were correct. We also present new high resolution $J$, $H$ and $K_{s}$ images,
  taken at Las Campanas Observatory (Chile).  All these data allow  to better investigate   the existence
  of detailed metallicity signatures
  on the dust emission which do not show up in the broad band emission 
  and to relate together dust, ionized gas and stellar content. These  data provide high 
  resolution analogs for the study of the ISM in conditions similar to those of  young galaxies 
  in the distant Universe which are becoming   more and more available  
  with the advent of large  ground based telescopes  (VLT, Keck, ALMA)  and space missions  
 (SPITZER, HERSCHEL).

\begin{figure*}
\includegraphics[width=0.9\textwidth,height=0.7\textwidth,angle=0]{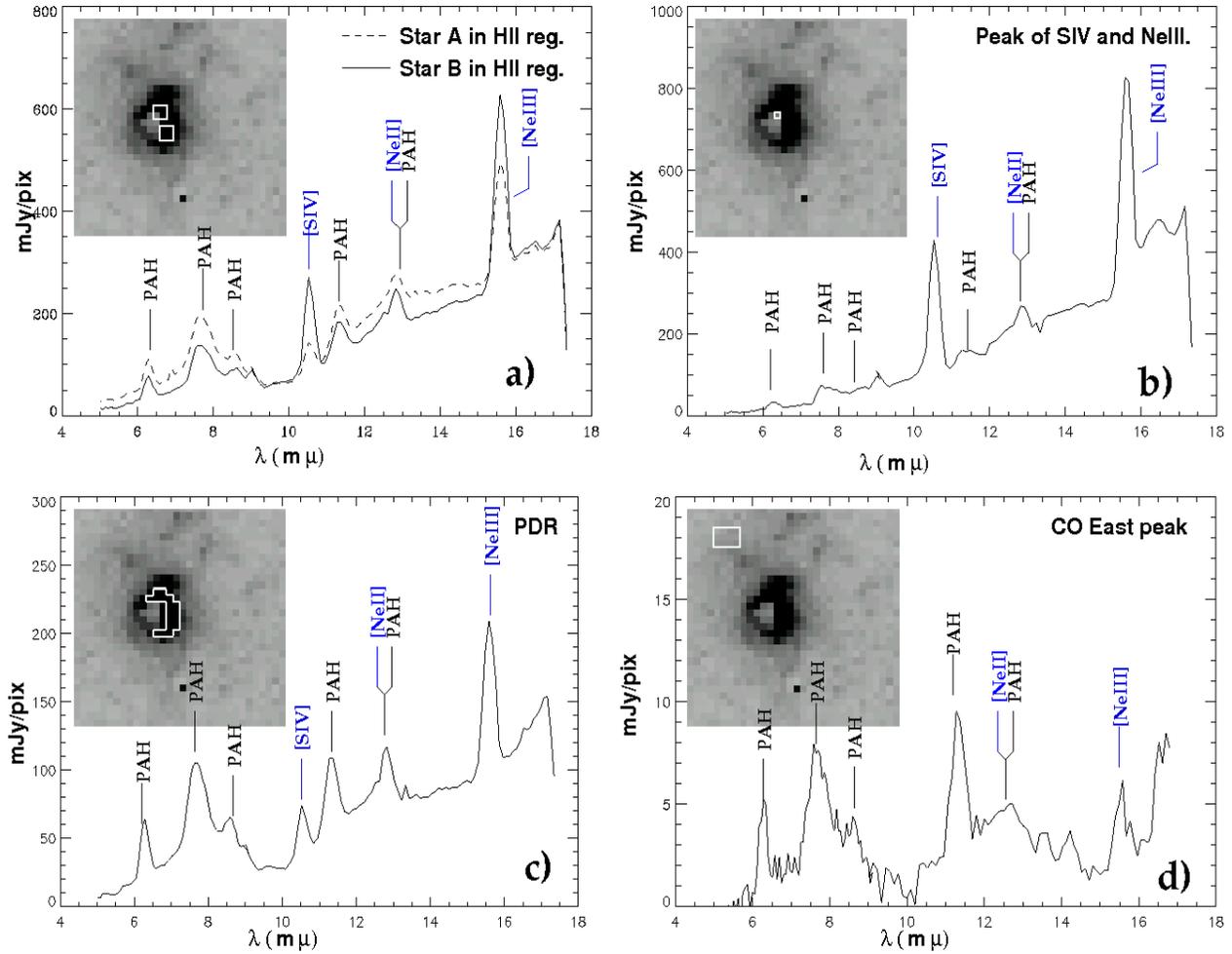}
\caption{ISOCAM CVF spectra of four different regions  of N4A. The regions where
the spectra were extracted are shown in each panel on the gray scale
image in the 7.7 $\mu$m aromatic feature. Note that the steep rise
or decrease at the very end of the spectrum are instrument artifacts and should not be considered as real.}
\label{Figure2}
\end{figure*}

\section{The data}
\subsection{ISOCAM-CVF data redution}

A complete scan of the two long-wavelength Circular Variable
Filters (CVF) was performed going down in wavelengths, step by step, first
from 16.2 to $9.0 \mu m$ (LW-CVF2) and second from 9.4 to
5.0 $\mu $m (LW-CVF1).
At each CVF step, between 7 and 11  frames of 2.1 sec were taken. The 
total observing time was 2974 sec. The pixel field of view was
6$^{\prime\prime}$ and the full field of the observation $3'\times 3'$ centered on RA=4:52:8.2 and DEC=
-66:55:16 (J2000).
The data reduction was performed as described in Boulanger \etal (\cite{BoulangerCVF}).
The reduced data cube can be retrieved from the ISO archive at the site
http://www.iso.vilspa.esa.es as Highly Porcessed Data Products (HPDP)
under the name Mid-IR Spectro Imaging ISOCAM CVF Observations.
 
\subsection{Near Infrared Broad band filters data}

Near Infrared $J, H$ and $  K_{s}$ deep images of N4A were obtained with the Dupont 2.5m 
telescope at Las Campanas Observatory (Chile) in the night of December 27, 1996  with the 256 x 256 NICMOS III camera IRCAM (Persson
et al. 1992). The spatial resolution of the system was 0.35" /pixel, and
the typical seeing was 0.9 \arcsec in $  K_{s}$.

The observation consisted of a series of 10 frames, each individual frame with
an integration time of 20 sec per filter. This procedure was repeated in a 9 
position mosaic with separation of 20 \arcsec. The total integration time in each filter resulted in
1800 sec in $  K_{s}$, 1800 sec in $H$, and 2000 sec in $J$. Sky frames in each filter 
were taken in  a field with faint stars and no extended 
emission at 12\arcsec W and 134\arcsec S of the source position. The sky field  
was observed in a similar way as N4. To produce the final images, each image
was flat fielded and sky subtracted, and then median averaged combined 
using IRAF procedures. The final images were registered with respect to the 
$  K_{s}$ image by means of several common stars. The resulting mosaic covers a 
65\arcsec $\times$ 76\arcsec area.

Aperture photometry in each
filter was performed with IRAF/DAOPHOT on the mosaic. For the purpose of this study we concentrated on the sources brighter
than $  K_{s}$ $< $13. The stars associated to N4A are shown in Figure 5. Table 1 gives the NIR 
magnitudes and colors of these stars. The photometric errors are $\sigma_Ks$ =0.06 mag,
$\sigma_H$ = 0.05 and $\sigma_J$ =0.03.  
Photometry of stars in our mosaic far from the N4A nebulosity were compared 
to their 2MASS photometry showing a good agreement. We estimated the
flux contribution from the nebulosity to our measured photometry  
to be less than 10\%  and 5\% in the
in the $K_s$  $J$ bands, respectively.

\section{Results}
\subsection{The Mid--Infrared Spectrum of N4A}
 Figure \ref{Figure2} shows the MIR spectra of 4 regions in N4A representative of 4 
typical components of an HII complex.
The upper left panel spectra (a) were produced by averaging 4
pixels (6$\arcsec$/pix) centered in the position of the 
two main exciting stars in N4A. The upper right panel (b) shows the spectrum
of the ionized gas peak emission.
The bottom left spectrum (c) is the average of  18
pixels on the bright north-east front of N4A; the bottom right  spectrum (d) corresponds to the CO eastern peak, 
outside the ionized cloud (Contursi \etal \cite{Contursi}).

 The spectra around the stars (a) and on the north--east front (c),
 are dominated by the AFEs and a continuum longward of 10 $\mu$m which 
becomes more and more prominent approaching  the stars (b and a). AFEs are stronger
 in the north-east front of N4A, as expected in a
region where both density and ISRF are high. This spectrum (c) is typical of PDRs confirming that this region is
 the interface between the HII region and the molecular cloud of N4A.\\
In the spectra  (panel a) corresponding to the regions close to the stars  the prominent
features are the fine structure gas lines of 
SIV at 10.5 $\mu$m, NeII at
12.8 $\mu$m  (although blended with the AFE at 12.7 $\mu$m) and NeIII at 15.6 $\mu$m. The AFEs over continuum ratio reaches its
minimum in the region where  the fine structure line are the most  intense (b) and AFEs practically
disappear. 
The  spectrum (d) towards the eastern CO peak  presents very weak AFEs and almost no continuum. 
These are characteristics of  quiescent regions and  suggest that the emission arises 
  from a thin external shell of the molecular cloud probably corresponding
to its red velocity component.\\ 
From this global spectral analysis  it appears that  N4  MIR dust and
gas properties are very similar to what is found in galactic HII complexes (Cesarsky \etal
\cite{M17}, \cite{N7023}), {\it i.e.} its MIR emission spectrum is typical of a 
HII/PDR/Molecular Cloud combination.

\subsection{Emission in the single dust bands and gas lines}
Thanks to the very high signal to noise and
the fact that we have spatial and spectral information on a region
of 3$\arcmin$$\times$3$\arcmin$ centered on N4A, this data set is 
perfectly suitable  for a detailed spectral and spatial analysis in the single
AFEs, gas lines and pure continuum.\\
We produced images in each dust feature (namely,
6.2 $\mu$m, 7.7 $\mu$m 8.6 $\mu$m 11.3 $\mu$m) and 
fine structure lines (SIV, NeII and NeIII) in the following way.
Following Boulanger \etal {\cite{Boulanger2}, we performed Lorentzian fit on the dust bands and Gaussian fit on the gas lines,
to the whole data cube, producing  
maps in each dust feature and   fine structure  line of the ionized gas.
First we fit together the 6.2 $\mu$m, 7.7 $\mu$m and 8.6 $\mu$m bands and a straight 
line as continuum letting the slope vary at each pixel.
Then, we fit together the SIV, 11.3 NeII(+12.7 $\mu$m) and NeIII lines, and a straight line as
continuum.
The reason why  we separately  fit the two
parts of the spectrum is because in the HII region core the continuum shortward and longward  10 $\mu$m 
might not have the same origin and therefore   the same slopes (see discussion in Sec.4.2). 
We then subtracted the fitted lines and bands from the original data-cube, pixel by pixel
producing  `pure' continuum spectra. We visually inspected the subtracted data cube
 to ensure that no `absorption' features due to overestimated lines/features were produced.
From the `pure' continuum spectrum,
we built images in the   ISOCAM  LW5 (6.5--7.0 $\mu$m) and  LW9 (14--16$\mu$m) filters.
They   represent the pure  continua shortward and longward 10 $\mu$m, respectively.
The NeII(12.7 $\mu$m) line was fitted with a Gaussian on the whole cube, 
even though we know that it is blended with the aromatic feature at 12.7 $\mu$m 
which should be fitted with a Lorentzian. Therefore the  NeII image contains also the 12.7
$\mu$m AFE.
The resulting images are shown in Figures \ref{Figure3} and \ref{Figure4}. 
All emission shown hereafter are
clipped at 3$\sigma$.\\

Figure \ref{Figure3}\footnote{Maps obtained 
from the CVF data have not  been  rotated
north-south because at these wavelengths the images are undersampled and a rotation would  conserve neither 
the flux nor the 
spatial distribution. To make the comparison among data at other wavelengths easier, we have rotated the
H$\alpha$ and the NIR images to the same ISOCAM orientation. }  shows  the 7.7 $\mu$m emission  obtained through Lorentzian
fits. Contours represent  
the SIV line emission  obtained applying Gaussian fits (gray) and the
  `pure' 15 $\mu$m (LW9) continuum emission (black).
In this image it is clear that the AFE   is coming from a shell surrounding the HII region
core and that this shell is particular bright in the PDR ({\it i.e.} in the North-East front). 
There are 2 bright compact knots on this shell (numbered in Figure \ref{Figure3}),  not resolved at the 7.7$\mu$m resolution
($\sim$7$\arcsec$). Most of the ionized gas emission arises  from the cavity formed by the dust shell
and it seems centered on a point like source. A similar result has been
published recently by Zavagno \etal (\cite{Zavagno}) on the massive galactic 
star forming region RCW 79. The 15 $\mu$m pure continuum
emission peaks close to knot\#2  in the PDR dust shell and 
has a weaker emission in the 7.7 $\mu$m cavity where SIV peaks. 
Figure \ref{Figure4} shows the 7.7 $\mu$m image with the NeIII contours. The general picture is 
very similar to that shown in 
Figure \ref{Figure3}:   the ionized gas traced by the NeIII emission
is mostly contained in the dust cavity and it does not show a secondary peak corresponding to
the secondary peak seen in the
 SIV line emission, although this can be due
in part to the poorer resolution at 15.6 $\mu$m than at 10.5 $\mu$m. The main
difference between the two ionized gas lines   emission is that the NeIII
emission  is much more extended than the SIV emission towards the south part of the nebula.
This is also the side where the HII region is partially bounded 
(Heydari-Malayeri and Lecavelier des Etangs \cite{Heydari}) and where the gas can expand. \\
All other  images in the other  dust bands  (not shown here) are very similar 
to the 7.7$\mu$m emission. The NeII(+12.7 $\mu$m) distribution is also very similar to
the distribution of the other AFEs everywhere but in the HII region core where 
there is a peak. This suggests, as expected, an increasing contribution of the NeII line to the
12.7 $\mu$m feature when approaching the exciting stars. We will come back to the
relative contribution of the NeII line and the 12.7 
$\mu$m AFE in Sec. 4.1.\\

 \begin{figure}
\includegraphics[width=0.4\textwidth,height=0.4\textwidth,angle=0,origin=c]{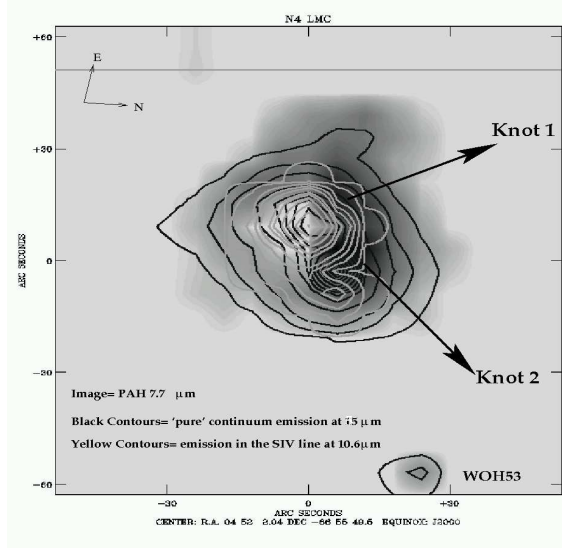}
\caption{Image: map from Lorentzian fit of the 7.7 $\mu$m dust emission feature.
Gray and
black contours: from
map of Gaussian fit in the SIV line and `pure' continuum in the LW9 filter
(15$\mu$m) emission respectively.}
\label{Figure3}
\end{figure}

 \begin{figure}
\includegraphics[width=0.4\textwidth,height=0.4\textwidth,angle=0,origin=c]{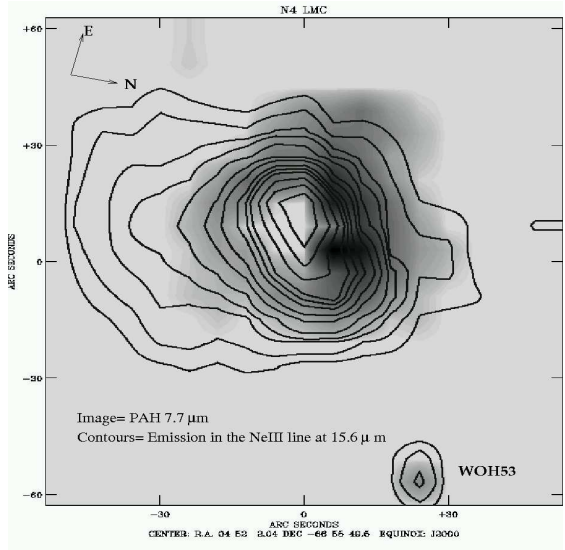}
\caption{Image: map from Lorentzian fit of the 7.7 $\mu$m dust emission feature. Contours: from
map of Gaussian fit in the NeIII line.}
\label{Figure4}
\end{figure}

\subsection{J,H and K$_s$ photometry.}
The $J-H-K_{s}$
composite image of N4A (Figure \ref{Figure5}) shows an extended and diffuse
emission whose morphology is very similar to the shape of the
H$\alpha$ emission seen in Figure \ref{Figure1} and  with  NIR color quite uniform 
across the nebula. 
Seven stars are visible (numbered in Figure \ref{Figure5}.).
Stars $\#1$ and $\#3$ are the bluest while the others are much redder.
Star $\#4$, at the north-west of the nebula, is bright and red. Star $\#2$ is located
at the border north of the nebula and shows diffuse emission around it.
%Photometry of all the seven stars was done and this is listed in Table 1.
Photometry of all the seven stars is listed in Table 1 ordered by $J$ magnitudes .\\ 
The comparison with the position of the stars indicated by Heydari-Malayeri 
and Lecavelier des Etangs (\cite{Heydari})   as the two main exciting stars 
(marked as A and B in Figure \ref{Figure1})  suggests that they  correspond to   stars $\#$1 and $\#$3  
in Figure \ref{Figure5}.  This 
association however, is not certain because of the lack of astrometry in their data (their Fig. 2).
We believe that these are the exciting stars because if one associates 
the H$\alpha$ knot visible in their Figure at about (0$\arcsec$, -27$\arcsec$) from
the center of their field with our star$\#$7, the position of the two
exciting stars as indicated in their paper corresponds to the position of stars $\#$1 and $\#$3. \\
The comparison between the $K_s$ band   and  the
15 $\mu$m `pure' continuum emission (shown in Figure \ref{Figure6}) clearly indicates
that  star$ \#$4  corresponds to the strongest peak of the pure continuum emission at 15 $\mu$m. 
Figure \ref{Figure7} shows the $K_s$ band emission contours on the SIV line image.
The SIV emission is almost entirely arising in the region where the  2 bluest stars are located.
Finally,  Figure \ref{Figure8} shows that one of the  AFE peaks (knot $\#2$ in Figure \ref{Figure3}) is close 
but not totally corresponding to
the bright continuum source (star $\#$4) and that knot $\#1 $ is located north of the blue stars, very close 
to where is the red elongated feature visible in the NIR composite 
map (associated with $\#$2 in Figure \ref{Figure5}).
This suggests that this source and star $\#$4 are embedded in a high density dusty region. \\
Note however, that the  uncertainty on ISO-CVF data   astrometry
($\leq$6$\arcsec$) plus
the fact that the images at different wavelengths are slightly displaced with
respect to each
other (by up 1 pixel over the whole ISOCAM spectral range)
prevents us to use the ISOCAM astrometry to superpose images at different
wavelengths.
The method we used to align  the CVF images with respect to each other and the CVF images on to 
the $K_s$ band image, is matching the peak emission  of the WOH53 star (marked in Figure 1).  \\
From the comparison between the ISOCAM and NIR data we conclude that SIV and NeIII emission, {\it i.e.} the ionized gas,
are mostly associated with the bluest stars in the cavity of the dust shell. The 15 $\mu$m continuum  peak
is mostly associated with the red star$\#$4 very close but not coincident to one of the peaks of
the   7.7 $\mu$m emission (knot $\#2$). Object $\#2$ corresponds to knot $\#1$ in the 7.7 $\mu$m emission.

 %NOTE: this is the old numbering already ordered by magnitude
%\begin{table}
%\begin{tabular}{ccccccc}
%\hline
%Star Nb. &   J$_{mag}$   & (J-H)     &  (H-K$_s$)   & (J-H)$_0$     &  (H-K$_s$)$_0$ \\
%\noalign{\smallskip}
%\hline
%\noalign{\smallskip}
%Star 2(A) &   13.63	   &   0.12	 & 0.24    & 0.07	 & 0.21   \\
%Star 7    &   14.16	   &   0.32      & 0.97	   & TO DO	 & To DO   \\  
%Star 3(B) &   14.34	   &   0.07      & 0.12	   & 0.03        & 0.08    \\
%Star 1    &   14.50	   &   1.01	 & 1.26    & 0.96	 & 1.23   \\
%Star 5    &   16.25	   &   0.93	 & 0.67    & 0.89	 & 0.63    \\
%Star 6    &   16.26	   &   1.01	 & 0.71    & 0.97	 & 0.67   \\ 
%Star 4    &   16.35	   &   1.04	 & 1.60    & 1.00	 & 1.56    \\
% \hline
%\end{tabular}
%\noalign{\smallskip}
%\caption{Observed and extincted corrected colors of the 6 stars stars visible in NIR images.  
%Numbers refer to Figure 5. Stars \#1 and \#3 are associated with stars A and B in figure 1.}
%\end{table}

%New numbering:

%      old        new

%star   2         1
%star   7         2
%star   3         3
%star   1         4
%star   5         5
%star   6         6
%star   4         7 

\begin{table}
\begin{tabular}{ccccccc}
\hline
Star Nb. &   J$_{mag}$   & (J-H)     &  (H-K$_s$)   \\
%\noalign{\smallskip}
\hline
%\noalign{\smallskip}
Star 1(A) &   13.63	   &   0.12	 & 0.24     \\
Star 2    &   14.16	   &   0.32      & 0.97	     \\  
Star 3(B) &   14.34	   &   0.07      & 0.12	     \\
Star 4    &   14.50	   &   1.01	 & 1.26     \\
Star 5    &   16.25	   &   0.93	 & 0.67      \\
Star 6    &   16.26	   &   1.01	 & 0.71     \\ 
Star 7    &   16.35	   &   1.04	 & 1.60      \\
 \hline
\end{tabular}
%\noalign{\smallskip}
\caption{Observed  colors of the 7 stars stars visible in NIR images.  
Numbers refer to Figure 5. Stars \#1 and \#3 are associated with stars A and B in Figure 1.}
\end{table}

\begin{figure}
\includegraphics[width=0.4\textwidth,height=0.4\textwidth,origin=c]{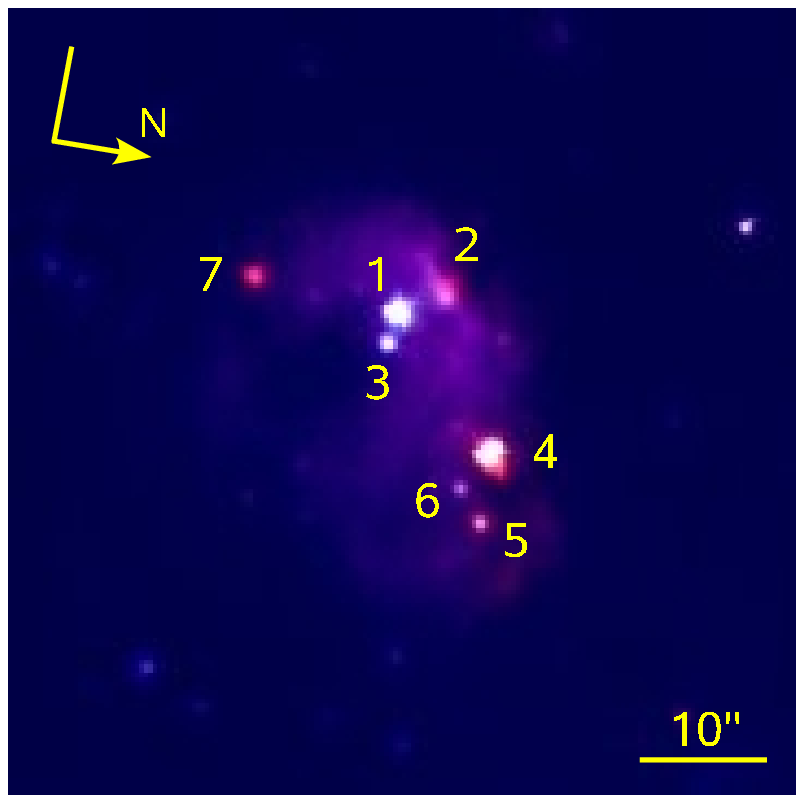}
\caption{NIR Composite map oriented as the ISOCAM images: red is $K_{s}$, green is $H$ and blue is $J$. Point sources identified as stars are
numbered as in Table 1. }
%NOTE FOR ME: This is the figure done by Rodolfo with the stars numbered according to
%their magnitude and oriented as the CAM images  
\label{Figure5}
\end{figure}

\begin{figure}
\includegraphics[width=0.4\textwidth,height=0.40\textwidth,origin=c,angle=0]{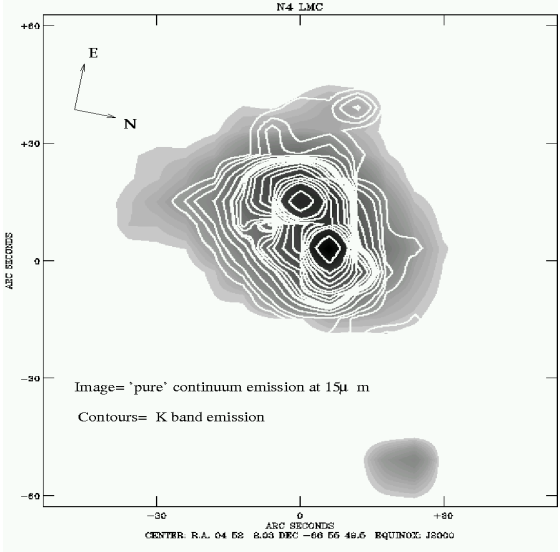}
\caption{Image: LW9 (15 $\mu$m)  pure continuum emission; Contours=   K band.}
\label{Figure6}
\end{figure}

\begin{figure}
\includegraphics[width=0.4\textwidth,height=0.40\textwidth,origin=c,angle=0]{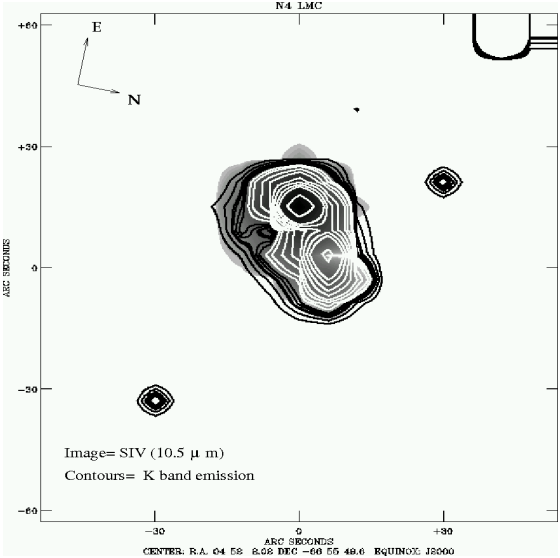}
\caption{Image: SIV at 10.5 $\mu$m from Gaussian fit; Contours=  K band.
Contours are shown in two colors (black and white) for visualization purpose.}
\label{Figure7}
\end{figure}

\begin{figure}
\includegraphics[width=0.4\textwidth,height=0.40\textwidth,origin=c,angle=0]{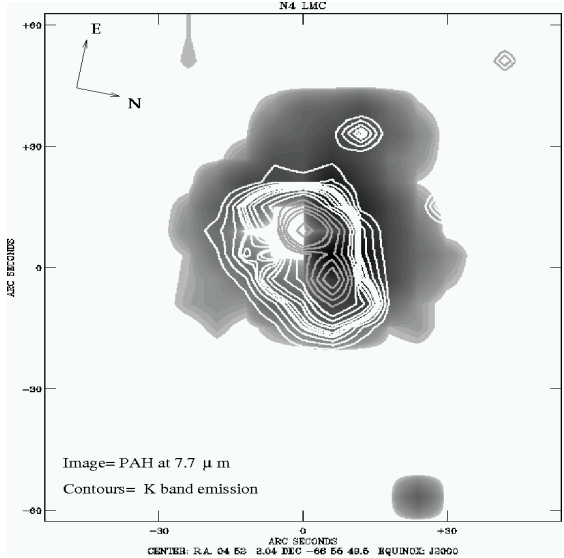}
\caption{Image: PAH, 7.7 $\mu$m feature from Lorentzian fit; Contours=  K band.}
\label{Figure8}
\end{figure}

\section{Discussion}
\subsection{The ionized gas}
Figure \ref{Figure9} shows the NeIII/NeII ratio contours over the 7.7 $\mu$m map.
As explained above,  this ratio has to be considered a
lower limit, because   AFE at 12.7 $\mu$m and the NeII at 12.8 $\mu$m are blended at
the ISO spectral resolution. 
In general, the infrared NeIII/NeII ratio, practically not affected by extinction (Thornley \etal \cite{Thornley}
and references therein),   is sensitive to the hardness of the UV radiation field. 
Thornley \etal \cite{Thornley} studied how this ratio changes in
a sample of star-forming regions belonging to starburst galaxies. They found that,
excluding the two systems (  IZw 40 and NGC5253) in their sample with low   metallicity, the
NeIII/NeII ranges from 0.05 to 1. 
In a comparable range of NeIII/NeII ratios, Brandl \etal (\cite{Brandl})
have shown that in a sample of starburst galaxies the Equivalent Width (E.W.) of the 7.7 $\mu$m  feature
does not vary significantly. At higher ($> 1$) NeIII/NeII ratios, 
a weak  anti--correlation   has been found by Wu \etal (\cite{Wu}) in a sample of Blue Compact Dwarf galaxies,
{\it i.e. low metallicity systems.}

In N4 the NeIII/NeII ratio peaks   in the HII region core, right on top of the bluest stars,
with a value equal to $\sim$10 which is comparable   to what  has been  found 
in IZw 40 and in 30Dor (Thornley \etal \cite{Thornley}) and in other low metallicity galaxies
(Wu \etal \cite{Wu}, Madden \etal \cite{Madden}).  
However, the global ratio is $\gtrsim$ 0.9, comparable with the typical value of the normal
 metallicity extragalactic HII regions. 
It is also noteworthy that there is an extended emission in NeIII/NeII(+12.7
$\mu$m) ratio map in the south part of the HII region. 

Table 2 shows the total flux of the fine structure  lines SIV, NeII(+12.7 $\mu$m) and NeIII. 
We checked whether the  emission in the ionized gas lines of N4A agrees
with the prediction of HII region models. We did not perform detailed calculation
with models   such as CLOUDY (Ferland \etal \cite{Ferland})  or similar, because we have only 3 lines available, one
of which is an upper limit.
We instead compared our results with the model prediction from Stasi\'nska
(\cite{Stasinska}) which gives the intensity of these lines relative to I(H$\beta$).
 Heydari-Malayeri and 
Lecavelier des Etangs (\cite{Heydari}) give an I(H$\beta$) intensity equal to
4.5$\times$10$^{-11}$ erg/s/cm$^2$ for a region centered on the exciting stars, of a
radius equal to 21$\arcsec$. In the same region we calculate the following
ratios:
  
\begin{eqnarray}
& &
\frac{I(SIV)}{I(H\beta )} = 0.34 \\\nonumber
& &
\frac{I(NeIII)}{I(H\beta )} = 0.6 \\\nonumber
& &
\frac{I(NeII)}{I(H\beta )} \leq 0.34
\end{eqnarray}

Assuming for N4 the physical parameters given in Heydari-Malayeri and 
Lecavelier des Etangs (\cite{Heydari} their tables 6 and 8), {\it i.e.}
 T$_{eff}$$\sim$40000-45000 K, $n$$\sim$100 cm$^{-3}$, [N]/[H ]=11.3e-6,
 [O]/[H]=2.6e-4, [Ne]/[H]=3.2e-5, [S]/[H]=6e-6, and a metallicity equal to Z/Z$_{sol}$$\sim$0.5,
  we found  that  the corresponding model from
 Stasi\'nska (\cite{Stasinska}) (IDD2) predicts the following ratios:
 
\begin{eqnarray}
& &
\frac{I(SIV)}{I(H\beta )} = 0.4 \\\nonumber
& &
\frac{I(NeIII)}{I(H\beta )} = 0.6 \\\nonumber
& &
\frac{I(NeII)}{I(H\beta )} = 0.05
\end{eqnarray}
 
  These  values compare very well with those observed in N4, listed in Eq. 1.
The model also predicts an   HII region radius of 4.2 pc, which corresponds 
   to an angular radius of 16.5$\arcsec$ (assuming a
 distance for LMC equal to  52 Kpc). This is quite the size
 of the NeIII/NeII emission if one does not include the extended emission on the south,
 outside the dust shell. Therefore, what we have called the {\it  HII region core}, finds here its
 justification. It also roughly corresponds to the size of  aromatic features cavity.

\begin{table*}
\centering
\begin{tabular}{ccccc}
\hline
Aperture  & F$_{10.5 \mu m}$(SIV)   &   F$_{12.8 \mu m}$(NeII+12.7)  & F$_{15.6 \mu m}$(NeIII)  &  \\
         &   10$^{-19}$~ W/cm$^2$  &     10$^{-19}$~W/cm$^2$        &  10$^{-19}$~W/cm$^2$ & \\
  \hline
Total &      68.7          &            51.6              &        65.0              & \\
This paper in ~SWS ~aperture &        8.3         &         6.8                  &         10.4  &\\       
Giveon et al. 2002 &        5.7         &         5.0                  &         13.7  &\\       
\hline
\end{tabular}
%\end{flushleft}
\caption{Flux in the fine structure lines from ionized gas:  first row gives the total fluxes; 
second row gives fluxes recovered in an aperture comparable to
that of ISO-SWS.}
\end{table*}

We now compare the flux we obtained with previously published similar  data from ISO-SWS.
Giveon \etal (\cite{Giveon}) studied the mid-infrared fine structure lines emission
of a sample of galactic and extragalactic HII regions, based on ISO--SWS spectra.
N4A belongs to this sample. 
In order to compare our flux with those given in Giveon \etal (\cite{Giveon}), 
we should extract the flux in the exact same 
aperture
as ISO-SWS (14$\arcsec$$\times$20$\arcsec$). This is not possible with 
the ISOCAM data due to the fact that the pixel size of the images is 6$\arcsec$
per pixel and any rebinning is unwarranted for quantitative calculation
because the pixel size undersamples the PSF at all ISOCAM wavelengths.
Therefore we can extract the flux in a box which is as close as possible
(12$\arcsec$$\times$18$\arcsec$) to the size of the
ISO-SWS aperture centered on the HII region peak. The fluxes we recovered are listed in Table 2.
They  agree within 30\% with the ISO-SWS fluxes. This is less than the total
uncertainties, calculated taking into account the ISOCAM photometry
uncertainties ($\simeq$10\%), uncertainties introduced by the fitting which we
estimate to be $\simeq$10\% and the ISO-SWS photometric uncertainties ($\simeq$30\%).
Thus the total uncertainty is   $>$ 33\% because we did not include
in the calculation  the uncertainties due to the
differences in apertures and in pointings which  cannot be estimated.
  
The agreement between the    fluxes we recovered from the ISOCAM  data
treated with the continuum plus feature fitting technique and  the ISO-SWS
fluxes, makes us confident on the reliability of the analysis presented here.
An interesting point is that the ISOCAM NeII+12.7AFE flux accounts for all the flux
in the ISO-SWS spectrum for NeII alone. This suggests that the contribution of the
12.7 $\mu$m dust emission in the center of the HII region is negligible as it could have been guessed from the
weakness of  the other dust features. Therefore, in the HII region core, the NeIII/NeII value
we calculated can be considered not an upper limit but the actual value which can now
be directly compared with  the  typical  ratios of  poor metallicity systems 
given by  Thornley \etal (\cite{Thornley}).\\
 
\begin{figure}
\includegraphics[width=0.4\textwidth,height=0.4\textwidth,angle=-90,origin=c,angle=90]{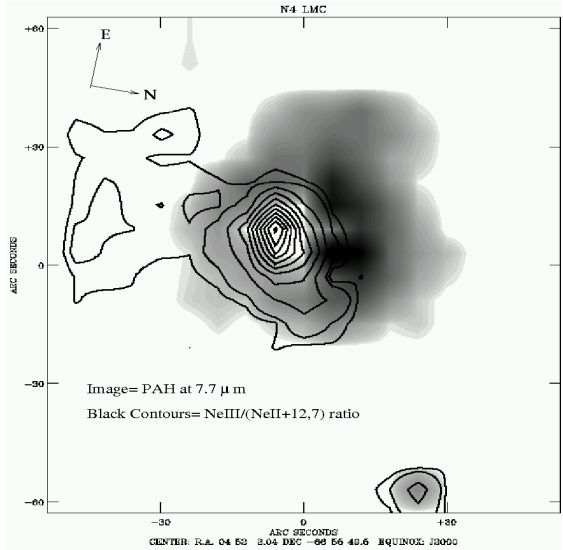}
\caption{Image: PAH, 7.7 $\mu$m feature from Lorentian fit; Contours=
NeIII/NeII(+12.7) ratio. Contour values are: 0.7, 1, 2, 3, 4, 5, 6, 7, 8, 9, 10.5. }
\label{Figure9}
\end{figure}

\subsection{Dust emission}
\subsubsection{Comparison with the current dust model}

  The AFE carriers are thought to be Polycyclic Aromatic Hydrocarbons (PAHs), composed by 50 up to few hundreds atoms,
and transiently heated mainly by Far UV photons (6--13.6 eV). Their emission intensity is proportional to the product
of the dust density and the ISRF, although it is certainly also depending on the hardness of the radiation field but in an
unknown way.
VSGs emit through a mechanism which is intermediate between the stochastic heating typical of  PAHs and the
thermal equilibrium of the classical big grains. In quiescent regions dominated by the general interstellar field, 
the dominant VSGs exciting mechanism is the stochastic heating which results in a  smooth continuum peaking between 30-40 $\mu$m. 
The emission mechanism becomes closer and
closer to a standard thermal behavior when the  radiation field gets more intense, such as in  star--forming regions and PDRs. 
In these conditions, because of their {\it quasi} thermal behavior, the VSGs emission spectrum shifts toward shorter
 wavelengths. 
 
Our observations  are in total agreement with the picture generally accepted to explain the dust feature and continuum
emission at MIR wavelengths. 
First, we have shown  in Sec. 3.2 that the morphology of the emissions in all dust features, including
the NeII+12.7 $\mu$m if one exlude the HII region core, are very similar to one another, which  strongly points towards a common origin and
excitation mechanism for all AFEs. 

%To investigate more this result we plot in Figure \ref{Figure10} upper panel
%the  the 7.7 $\mu$m {\it versus} the 8.6 $\mu$m dust emission features. This is
%simply a pixel to pixel plot after having rebinned the dust maps 
%by 2, in order to have a pixel size (12$\arcsec$/pix), {\it i.e.}  greater than
%the ISOCAM PSF (FHWM= 7$\arcsec$--9$\arcsec$) in the wavelengths range covered by 
%ISOCAM. The plot clearly shows that there exists a positive correlation 
%(correlation factor {\it r} = 0.8) between the AFEs.

Second, the spectra of N4 A presented in Figure \ref{Figure2}  can be fully explained with 
the classical picture illustrated above: 
in the HII region core the continuum longward
10 $\mu$m increases just because  VSGs reach the thermal equilibrium regime, get hotter, thus producing a
shift of their thermal emission spectrum  to   wavelengths shorter than 30 $\mu$m. The opposite happens
 in quiescent regions, such as
that corresponding to the eastern CO peak outside the HII region, where in fact almost 
no continuum is detected.

\subsubsection{Origin of the continuum at MIR wavelengths}
 Since we have built  `pure'
continuum images at 6.75 and 15 $\mu$m, we can also investigate how these continua relate to AFEs.
The origin of the MIR continuum in star-forming regions, PDRs and diffuse ISM, is a controversial issue.
In principle it can be due to hot VSGs which emission, as we have explained above, shifts into the MIR ISOCAM
wavelengths and/or a continuum related to the AFE (maybe simply due to a superposition of 
unresolved and weak bands). 
Figure \ref{Figure10} shows the 6.7  $\mu$m `pure' continuum' contours on the 15 $\mu$m `pure' 
continuum emission image smoothed at the same resolution. Maps have
been aligned on the WOH53 star. The two continua are very similar suggesting a common
origin. On the other hand their   morphology  is not totally similar to the dust
feature emission morphology:
they  are more centrally peaked than the dust feature emission which forms a cavity.
  We therefor conclude that,    at least in conditions similar to those of
N4, {\it i.e.} in HII region complexes, most of the continuum at MIR wavelengths , 
even at wavelengths shorter than $\sim$10 $\mu$m, arises from VSGs and not from unresolved 
weak aromatic bands. This might not be the case in more quiescent regions
like diffuse ISM, as indicated by the  spectrum corresponding to the eastern
CO peak (Figure \ref{Figure2}) where almost no continuum at all has been detected.

\begin{figure}
\includegraphics[width=0.4\textwidth,height=0.4\textwidth,angle=0,origin=c]{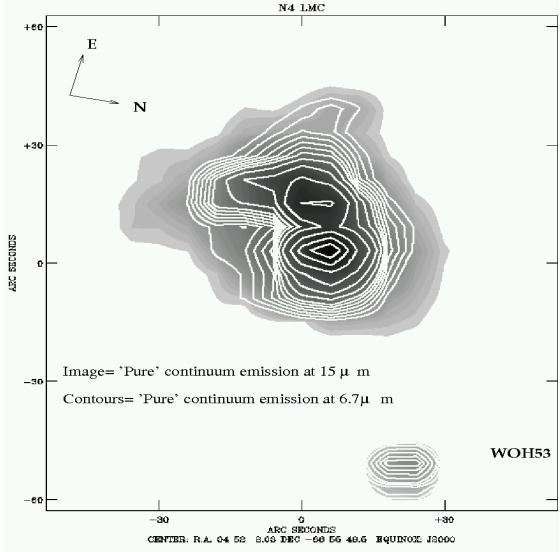}
\caption{Image: map from the 15 $\mu$m 'pure' continuum emission. Contours: `pure' continuum at
6.7 $\mu$m. smoothed to the 15  $\mu$m resolution.}
\label{Figure10}
\end{figure}

\subsubsection{Why is there  no AFE in the HII region core?}
In this section we investigate the reason why  no AFEs are detected in the HII region core.
The depression of AFE in high ISRF environments has been already
 observed in the Milky Way (Cesarsky \etal, \cite{M17},\cite{N7023}), Small Magellanic Cloud (Contursi \etal, \cite{N66}) and in
 external galaxies (Galliano \etal \cite{Galliano}, Lu \etal \cite{Lu}, Houck \etal \cite{Houck}). Recently, a
 high resolution study of the 3.3 $\mu$m AFE (not included in the ISOCAM--CVF wavelength range) with ISAAC at
 the VLT in NGC253 and NGC1808 (Tacconi--German \etal \cite{Tacconi}), has shown that AFE are strongly depressed with respect to
 the continuum  close to where the super star clusters are located. For this reason these authors concluded that the PAHs may be a better
 tracer of B stars rather than more massive and earlier spectral type stars,   as was already 
 previously suggested by Peeters \etal (\cite{Peeters}) on the basis of the analysis of a wide
 sample of galactic HII regions observed with ISO. 
 
In the case of N4, almost no AFE is detected at all in the HII region core. 
This can in principle be explained with three possibilities:
\begin{itemize}
\item{{\bf 1)} stellar winds are capable to evacuate the HII region
from AFEs carriers but not from gas;}   
\item{{\bf 2)} Radiation pressure is capable to push away the dust but not the gas;}
\item{{\bf 3)} AFEs carriers are destroyed in very extreme Radiation Fields (RF).}
\end{itemize}
 Figure \ref{Figure3} shows 
 that the `pure' continuum emission at 15 $\mu$m has a secondary peak in the region 
 where the ionized gas peaks. This is also visible in the MIR spectrum centered on 
 SIV and NeIII peak (Figure \ref{Figure1} panel b) where almost no AFEs are visible but a 
 significant continuum  is still detectable. This shows that dust can survive in HII cores.
 As a consequence, the first two possibilities would imply that  AFE carriers, VSGs and gas are 
 decoupled  which is   physically   quite improbable. 
 We therefor conclude that   the dominant mechanism has to be the destruction of AFE
 carriers.\\ 
The question now is  whether dust destruction is due
to  the intensity or the hardness of 
the radiation field, or both. 
Recent results from Wu \etal (\cite{Wu}) and Brandl \etal (\cite{Brandl}) have shown 
that PAHs E.W. stay quite constant for a wide range of UV hardness as traced by the NeIII/NeII
ratio and  start to slightly decrease only at high ($\gtrsim 1$) NeIII/NeII ratio in low
metallicity galaxies. More important is the result from Wu \etal (\cite{Wu}) 
 who showed that PAH E.W. is
strongly  anti correlated with the {\it product} of the radiation field intensity and hardness.

 We tried to investigate this issue as
 illustrated in Figure \ref{Figure11}. Here both panels
 show  the 7.7 $\mu$m dust feature emission, the 6.75 and 15 $\mu$m `pure' continua emission  
 as function of  the 
ISRF energy density at 1600 $\AA$ (left panel) and as function of the  NeIII/NeII ratio (right panel). The energy density at 1600 $\AA$  is in units of the solar neighborhood at the same
wavelengths as measured from Gondhalekar  Phillips and  Wilson
(\cite{Gondhalekar}) and it   has been
estimated in Contursi \etal (\cite{Contursi}). These authors calculated at each wavelength the average emissions  
in annuli centered on the main ionizing stars till 3 $\sigma$ level.
The NeIII/NeII ratio is a tracer of the UV hardness. However, as already discussed in Sec.
4.1, at the ISOCAM spectral resolution, the NeII line emission at 12.8 $\mu$m  is blended with the
AFE at 12.7 $\mu$m.
Therefor, in the case of N4, this ratio is a lower limit everywhere but in the HII region core,
where   most of the flux can be  ascribed to  the NeII emission 
line alone, as it has been shown in Sec.  4.1.
 In order to enhance differences among AFE and 'pure' continua emission, we have normalized each emissions
 to their maxima. \\
 Form Figure \ref{Figure11} we observe the following behaviors: 
\begin{itemize}
\item{All dust components (AFEs, and the continua at the two wavelengths)  
show the same behavior: their emission  increase till a certain
 value going from the outside  toward the
  HII region center.   They reach a maximum presumably on the PDR and then sharply
  decrease  in the HII region core.}
%We note however, that this change is much sharper 
%when the emission is represented as function 
%of the NeIII/NeII ratio suggesting  that
%there must be a UV hardness threshold value above which destruction becomes very efficient.\\
%
\item{The AFEs's drop is the steepest, followed by the  the 6.7 $\mu$m continuum and then
by the 15 $\mu$m continuum.}
\item{The AFE's drop occurs first, {\it i.e.} at ISRF density and UV hardness lower than
for the pure continua emission}
\item{ The decline of all emission appears steeper when plotted versus the NeIII/NeII ratio than
when plotted as function of the ISRF energy density}
\end{itemize}

These results definitely point  to a scenario where, given the same conditions,
AFE carriers are more easily destroyed than hot dust. In other words, these
carriers are destroyed by UV photons  less energetic than those necessary to destroy 
 the carriers responsible for the 'pure' continuum emission. If, following current
dust models, these two types of
carriers are associated with PAHs and VSGs,  we can interpret this
 behavior as an evidence that PAHs and   VSGs are both destroyed but with 
 different efficiencies. Moreover, assuming that the 
 VSGs population has  a certain size  distribution, we can reasonably
 think that 
 the continuum emission at lower wavelengths is dominated by smaller particles than those
 responsible for the emission at longer wavelengths. In this framework,
 the fact that the drop of the continuum at 6.7 $\mu$m is stepper than the drop of the 15 $\mu$m
 continuum, can be interpreted as the fact that the  destruction mechanism is more efficient 
 on smaller molecule/grain. This means that: {\it the destruction
 mechanism is likely to affect the dust size distribution}. \\
 This conclusion may have some important implications for understanding the ISM physic in dwarf galaxies.
 Thornley \etal (\cite{Thornley}) found
a NeIII/NeII ratio for IZw40 similar to what we measured in the very center of N4A (roughly a shell of
$\sim$10 pc in radius) but  in a much larger region ($\sim$700$\times$$\sim$1000 pc). 
This means that the phenomena of destruction in such galaxies must occur in 
much larger regions than in normal galaxies. This can be explained 
by the fact that dwarf low metallicity galaxies have an average UV spectrum harder
than normal or starburst galaxies, due to a higher star formation rate per unit mass and
to the presence of more young and massive star clusters. The UV photons can travel longer
before being absorbed thanks to the fact that these systems have less dust (Madden \etal
\cite{Madden}).
{\it All these factors point to a scenario where   
in dwarf galaxies, PAHs have a higher probability to be destroyed than in normal metallicity
environments leading to an overall change of the galaxy dust size distribution. It seems also that,
at least in not extremely poor metallicity systems,  this phenomenon is more important
than intrinsic chemical dust properties modification}.\\
Unfortunately, from this analysis 
 no firm conclusion can be  achieved about
 the possible different roles played by the ISRF intensity and hardness on the dust
 destruction mechanism. However,   a weak evidence that the latter 
 has an  impact stronger than the ISRF intensity  might be deduced by the steeper 
 drop all dust emission show with the NeIII/NeII ratio 
 than that shown with the ISRF energy density.

\begin{figure*}

\includegraphics[width=0.4\textwidth,height=0.8\textwidth,angle=-90,origin=c]{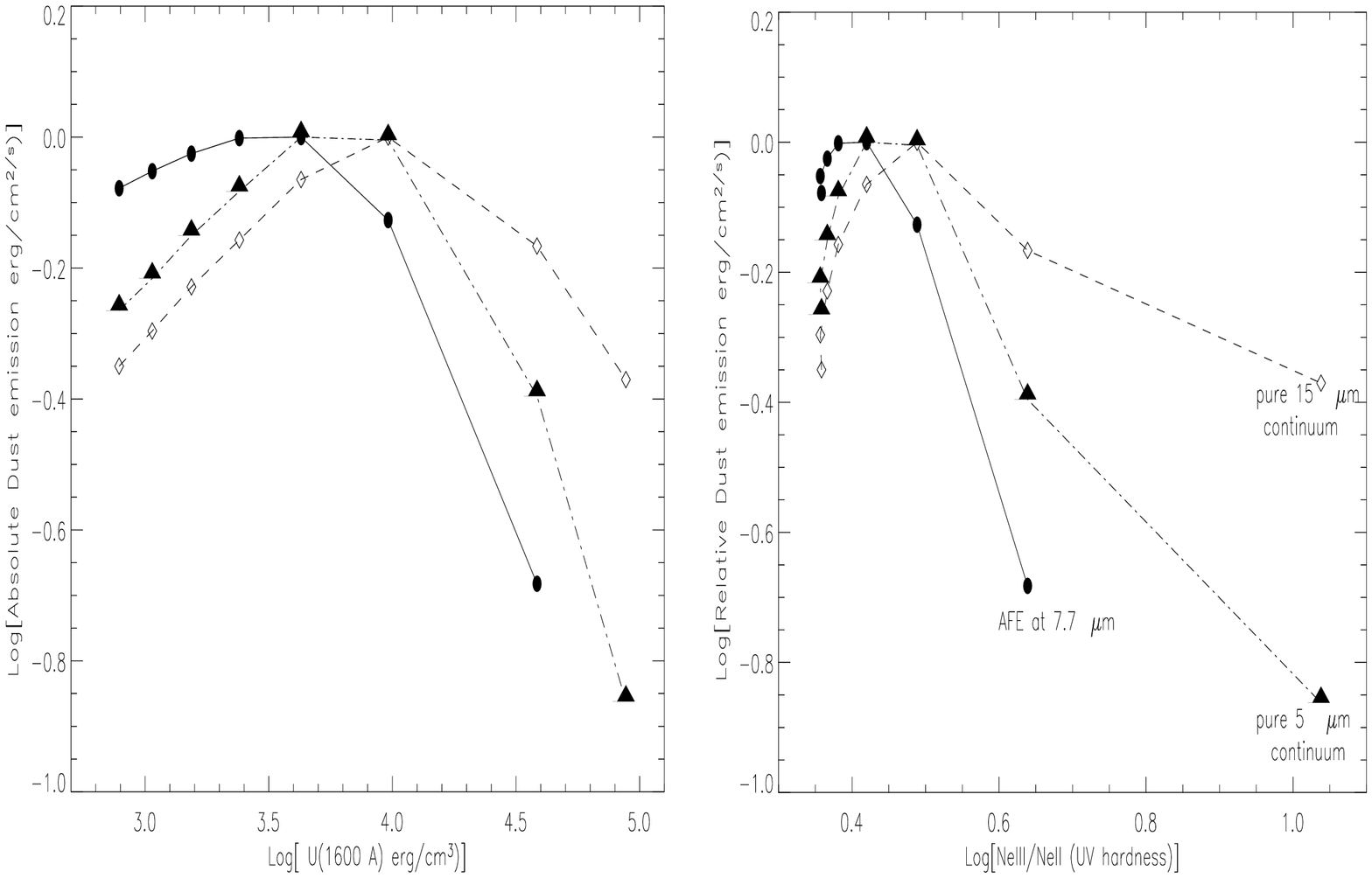}
\caption{The 7.7 $\mu$m PAH emission (filled circle) 5$\mu$m and  15 $\mu$m pure continuum
emission (triangles and diamonds respectively)
{\it versus} the intensity of the radiation field at 1600\AA~ (left panel) and NeIII/NeII ratio
 which traces the UV hardness (right panel). Values are above the 3$\sigma$ for the 3 wavelengths. In the right
 panel, the maximum at 7.7 $\mu$m and 5 $\mu$m have been scaled to the maxima at 15 $\mu$m to ease the
 relative comparison among the different emissions.}
\label{Figure11}
\end{figure*}

\subsection{Comparison with hot dust properties in galactic environments.}
 Metallicity is
a  parameter which  may   significantly affect dust. In low metallicity environments, 
such as those we expect to find in young distant galaxies, we expect   dust  to have  
intrinsic chemical-physical   properties  
very different from those observed in normal metallicity galaxies, 
just because dust forms and grows  in an ISM with 
very different
chemical enrichment and mechanical inputs (from both stellar winds and supernovae). These might
lead to un overall under-abundance of carbon based grains with respect to the abundance in normal
metallicity environments but also to dust feature emission profiles, widths and intensity ratios
different form those found in normal metallicity ISM.\\
 In   Section 4.2.3 we have shown that in N4 dust destruction  could be, at least in part, 
 responsible for lack of
small molecules/grains, already detected on much larger scale with IRAS 
in the Magellanic Clouds  (Sauvage, Vigroux and Thuan \cite{Sauvage}), although we might
have a total  amount of carbon based molecule/grains lower than in  Galactic HII regions. On the
other hand, we have detected strong aromatic features very similar to what is observed in the
Milky Way's ISM. In this section we will analyse in more details the N4 dust features widths and intensity
ratios  to understand their physical characteristics and how do they compare with
similar observations in galactic environments.\\

\begin{itemize}
\item{{\bf Dust feature intensities.}\\
Figure \ref{Figure12} shows a comparison between the PDR spectrum  in N4
 (panel c in Figure \ref{Figure2}) and
the spectrum of one pixel in the PDR of NGC7023, after having subtracted a straight line continuum 
from both and scaled up the spectrum of NGC7023 to match by eye the 7.7 $\mu$m feature of N4.
There are no evidences,
at least at the CVF   spectral resolution ($\lambda$/$\Delta$$\lambda$$\sim$50)
 of grain/molecules modifications due to
metallicity resulting in differences in the features profiles or peak wavelengths shifts. Indeed,
the shape and the relative intensities of the 6.2, 7.7 and 8.6 $\mu$m features in NGC7023 and N4 
match almost perfectly. There is an indication of a   relatively  lower 11.3 $\mu$m feature in N4 than in
NGC7023, which may be due to a higher level of PAH ionisation  in N4 . This is plausible, since N4
is more active than NGC7023, even in its PDR region as it is witnessed by the presence of ionized
gas  lines totally absent in the NGC7023 spectrum (but see more details below in
this Section).
This confirms the result obtained by Vermeij \etal (\cite{Vermeij}) which found that the AFEs  characteristics of  LMC HII regions are similar to those
in the Milky Way. }
\begin{figure*}

\includegraphics[width=0.8\textwidth,height=0.8\textwidth,origin=c]{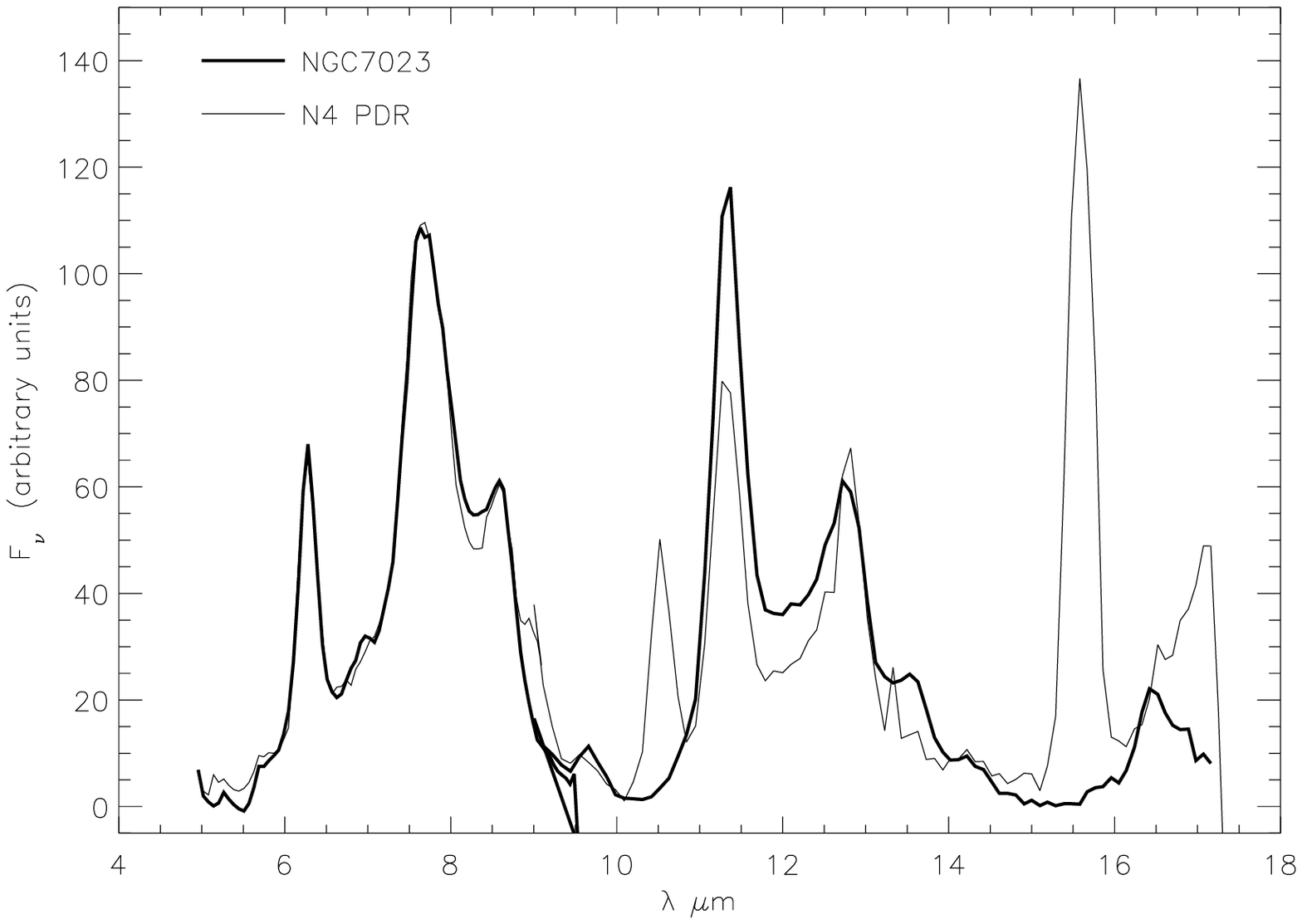}
\caption{Comparison between the spectrum of the galactic region NGC7023 (thick solid line) and the PDR spectrum of
N4 (thin solid line), after continuum subtracted. NGC7023 spectrum has been scaled to match the 7,7 $\mu$m feature
of N4.}
\label{Figure12}
\end{figure*}

 \item{{\bf Dust features widths}\\
The analysis of the dust features conducted in this work is based on the assumption that their profiles  
are represented by Lorentzian.
In this context,  the widths of the features have a physical explanation.
Boulanger \etal (\cite{Boulanger2}),  proposed  that the emitting mechanisms of the AFE's carriers 
is the internal vibrational redistribution. The bands
widths arise from intramolecular broadening processes in large molecules and  are related to the very short life--time of 
the emitting levels.\\
One of the byproducts of the fitting technique we have used, are the widths of the 
fitted Lorentzian through the entire data cube  ({\it i.e.} wavelengths).
We can therefore compare what obtained for N4 with the range of widths that Boulanger 
\etal (\cite{Boulanger2})  found for typical  galactic regions 
(namely $\rho$ Ophiuchus and NGC7023). 
Figure \ref{Figure13} shows such a comparison. We have calculated the mean  widths of N4 
obtained by the fit, taking into account all pixels with a signal greater than the 3 $\sigma$
of the {\it r.m.s.} at each wavelengths (filled circles) and the median for the PDR region only, 
{\it i.e.}  the same region shown in panel c of Figure \ref{Figure2} (filled triangles). The open stars 
 corresponds to the mean FWHM values found for the two galactic regions by Boulanger \etal 
 (\cite{Boulanger2}). We exclude the 12.7 $\mu$m feature because we fitted it with a Gaussian 
 and not with the Lorentzian as  in  Boulanger \etal {\cite{Boulanger2}.
 The agreement is very good at all wavelengths showing that also this parameter is very similar to
 that observed in galactic environments despite the fact that  $\rho$ Ophiuchus and NGC7023
 are in ISRF much less intense than  N4.}}\\

\item{{\bf Dust feature  ratios}\\
 Reach et al. (2000)  found that the 11.3 /7.7 $\mu$m ratio 
was higher in SMCB1-1, a relatively quiescent cloud in the Small Magellanic Cloud (SMC)
not associated with on-going star formation,
than in various ISM regions in the Milky Way. This was
attributed to an enhancement of the C-H bonds with respect to the C-C
bonds,
due to the higher H/C abundance ratio in the SMC than in the Milky Way. 

We have produced   AFE feature ratio maps 
(not shown here), to study their variation across the region. The mean 11.3/7.7 $\mu$m
(C--H/C--C) value is 0.25 very similar to what found by Boulanger et. al. (1996) in
$\rho$-Oph and smaller than the value in SMCB1-1 (~0.8).
A collection of 11.3/7.7 $\mu$m ratios in different galactic ISM by
Lu \etal (1998) though, clearly indicates that this ratio has a large dispersion 
even in similar  ISM phases. Therefore,  the analysis of the dust feature ratio
confirms that in N4 the dust  feature ratio are comparable to what found in many other regions.
An  important result  is that this ratio is quite constant on
the dust shell, indicating that whatever is the state of the AFE carriers,
{\it i.e. ionized/neutral, hydrogenated or modified from metallicity}, 
there are no strong differential effects on the dust feature emission, and therefore presumably
on the dust conditions, at least at 
this spatial resolution ($\sim$ 2 pc assuming a LMC distance of 52 Kpc).\\

The changes in the   profiles  of the dust features and their relative intensities can tell a lot
about the physical state of AFEs carriers if associated with PAHs
(Pauzat, Talbi and Ellinger \cite{Pauzat}, Le Page, Snow and Bierbaum \cite{Page}, Dartois and
D'Hendecourt \cite{Dartois}, Bakes, Tielens and Bauschlicher \cite{Bakes}). 
While the first analysis
is impossible at the ISOCAM-CVF spectral resolution, we can analyze how AFE
 ratios change in N4 as function of the ISRF  intensity. 

It  is believed that PAHs in PDRs are
mostly  ionized. Ionization of PAHs leads to an enhancement of the intensities
of the features arising from the C--C bonds (6.2 and 7.7 $\mu$m bands in the ISOCAM wavelengths
range) with respect to those arising from the C--H bonds (8.6 $\mu$m in--plane  and out-of
plane (oop) solo, duo, trio and quatro at  11.3, 11.9, 12.7 and 14 $\mu$m, respectively).
Dehydrogenation weakens the C--H bonds. Therefore the ratio
between any feature arising from C--H bonds with a feature arising from C--C bonds, should
decrease as ionization and dehydrogenation increase. 
Figure \ref{Figure14} upper panel shows the 11.3/7.7 $\mu$m ratio (oop C--H/
vibration C--C) as function of the ISRF. There is a weak decreasing trend indicating that
either ionization and or dehydrogenation are perhaps weakly increasing as approaching the HII region core.
On the other hand, in the very center there is an increase of this ratio  probably due to the
fact that here smaller molecules (which emit at shorter wavelengths) are destroyed before 
larger species. 
Verstraete \etal (\cite{Verstraete}) analyzed the   oop C--H  to in plane C--C dust features
ratio, traced by the (11-13)/(6.2+7.7) $\mu$m ratio. Theoretical predictions show that this ratio is
2.5 for neutral PAHs and 0.25 for cations. Figure \ref{Figure14} bottom panel shows a similar
ratio where in place of the 11-13 emission features we calculated the 11.3+12.8 $\mu$m
emission. These ratios are well below those predicted for neutral PAHs everywhere in  N4,
and very close to the value for ionized PAHs,
suggesting that  PAHs are ionized in the whole region. Note that in the HII region core, this 
ratio is an upper limit because the
12.7 $\mu$m emission is almost entirely due to NeII rather than the AFE at 12.7 $\mu$m. 
We therefore conclude that {\it if PAHs are the carriers responsible for the dust features,
in N4 they are almost totally ionized, and  that their state of  ionization and /or dehydrogenation is
quite constant all over the region}. Probably, in order to detect the transition between neutral to ionized PAH, one has trace the
PAH emission from diffuse ISM to active regions. }
\end{itemize}

  In conclusion, we have clearly shown that three fundamental parameters of AFE, namely, emission intensities, features
typical widths and emission ratios are all very similar to what is typically found in normal
metallicity ISM.

\begin{figure}
\includegraphics[width=0.45\textwidth,height=0.45\textwidth,origin=c]{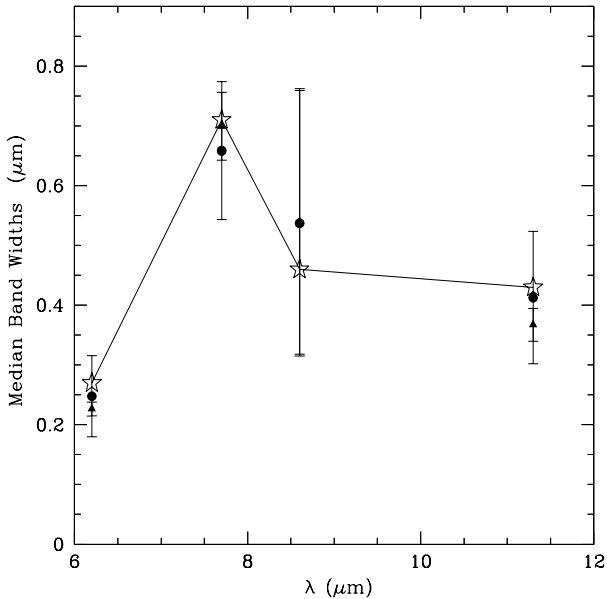}
\caption{Typical widths of the AFEs obtained by the Lorentzian fit for all N4  (filled circles),
the  PDR only (solid triangles) and the typical values found with the same fitting technique by
 Boulanger \etal (1998) in $\rho$ Oph and NGC7023 (open stars). }
\label{Figure13}
\end{figure}

\begin{figure}
\includegraphics[width=0.35\textwidth,height=0.45\textwidth,origin=c]{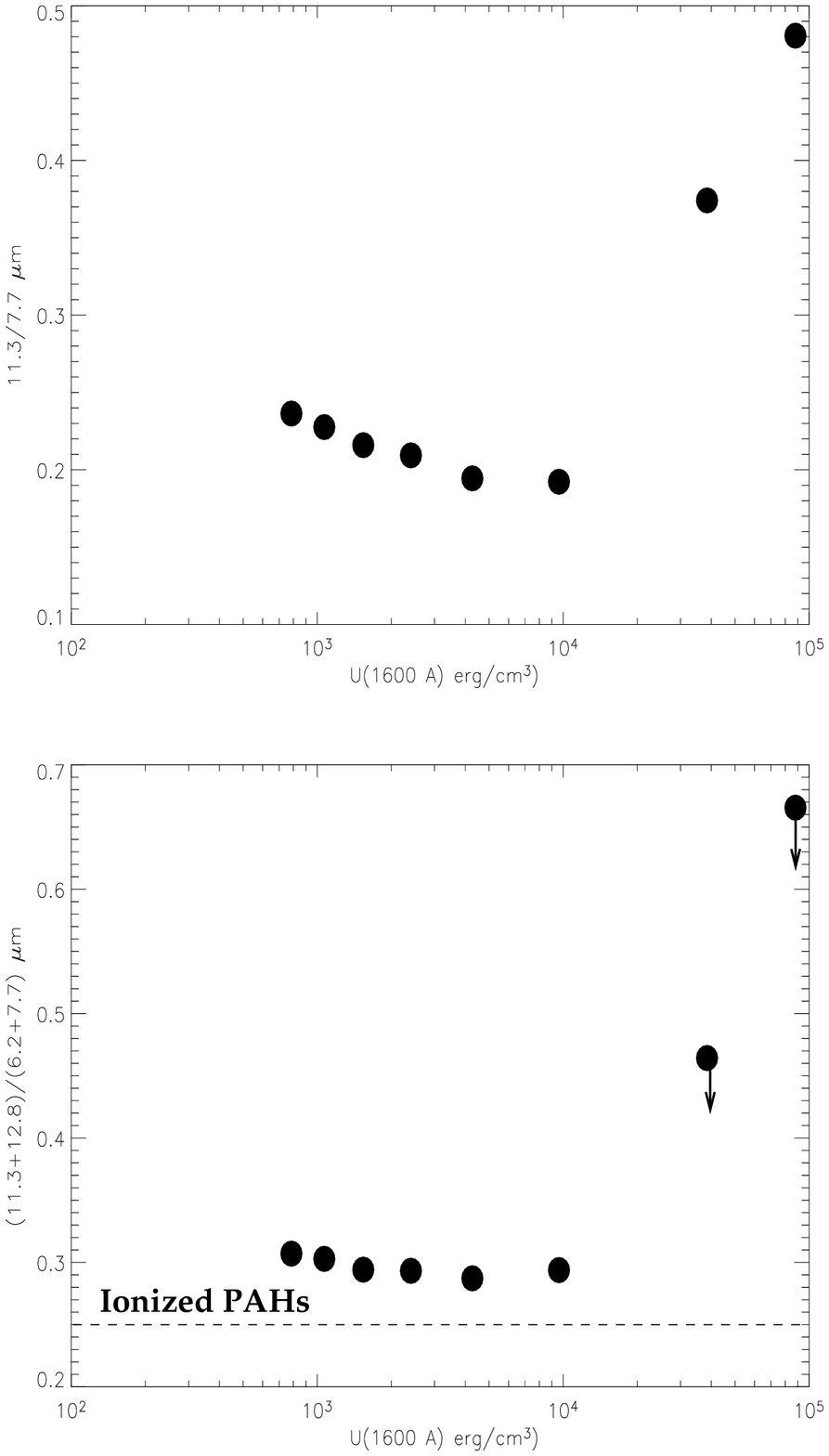}
\caption{Different dust feature ratios as function of the ISRF. Upper panel: 11.3/7.7 $\mu$m dust feature ratio, which traces the 
[out of plane C--H]/[vibration C--C] ratio. Bottom panel: (11-13)/(6.2+7.7)$\mu$m dust feature ratio which traces the 
[out of plane C--H]/[in plane C--C] dust features ratio.}
\label{Figure14}
\end{figure}

\newpage
\begin{figure*}
\includegraphics[width=0.50\textwidth,height=0.99\textwidth,angle=-90]{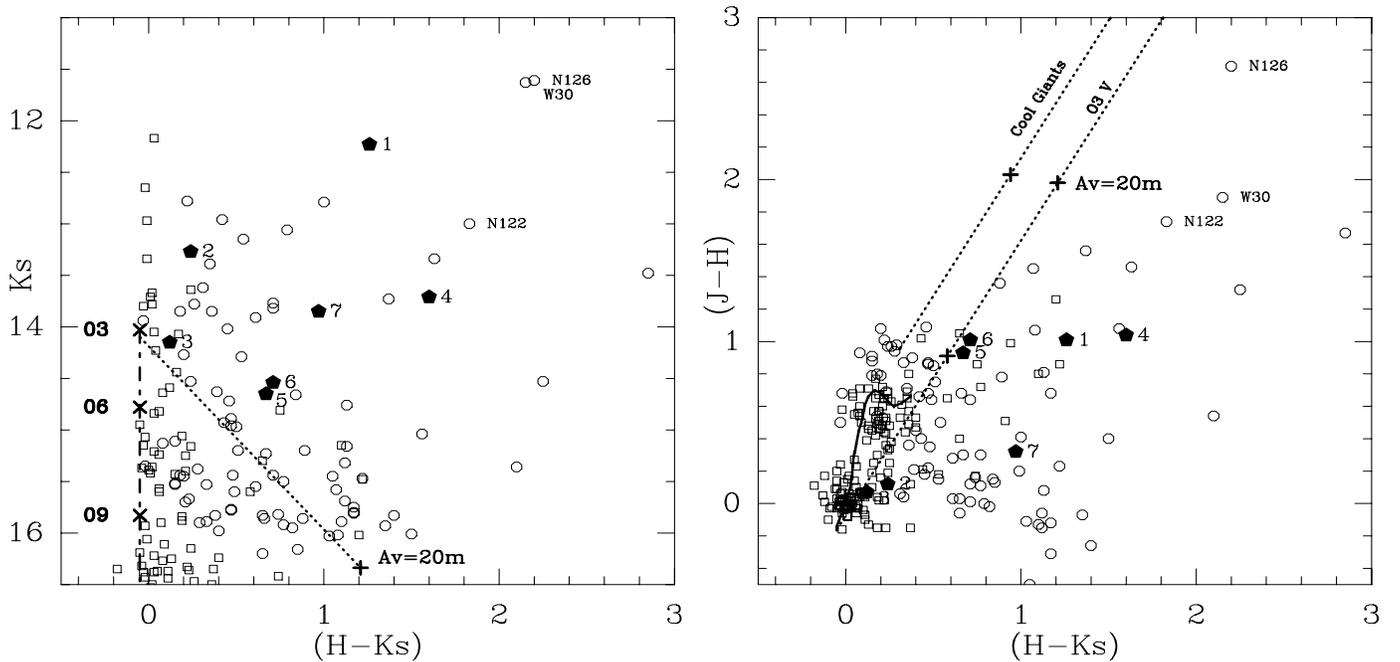}
\caption{NIR color-magnitude (left) and color-color (right) diagrams of
N4 with  the IR sources  showed as red filled points. 
For comparison we include the 30 Doradus IR sources as open 
circles  and the N11 IR sources as open squares.
In the color-magnitude diagram the upper zero-age main sequence
between O3~V and O9~V corresponds to a distance modulus of 18.6 and is
indicated with a dashed line.  The reddening track for a normal O3~V
star is plotted with a dotted line and extends to $A_V = 20$ mag. 
In the color-color diagram, the main-sequence locus from O3~V to M2~V and
the cool-giant branch are indicated by a solid line. The reddening tracks 
for normal O3 V and cool giant stars are plotted as dashed lines, 
with crosses indicating $A_V = 10$ and 20 mag. Sources 1, 4 and 7  show IR excesses which 
cannot be explain from reddening.}
\label{Figure15}
\end{figure*}

\section{The stellar content}
 
The NIR data presented in this paper unveil at least 7 bright sources.
Figure \ref{Figure15}  show a NIR color--magnitude  diagram (right panel) and $JHK_{s}$ color-color 
(left panel) of stars numbered as in Figure 5 whose photometry is reported
 in Table 1. These are shown as red pentagons in the Figure \ref{Figure15}. 
This Figure includes for comparison the IR data obtained for two other 
LMC star forming regions, 30 Doradus  (Rubio \etal \cite{Rubio}) 
and N11 (Barb\'a \etal \cite{Barba}). In each  figure's panel
is plotted the zero age main sequence (ZAMS) adopted from 
Hanson, Howard and Conti (\cite{Hanson}) and the extinction
tracks for the reddening law of Rieke and Lebofsky (\cite{Rieke}).

Note that the extinction for star $\#4$ is likely to be much higher
than the average as this star falls in the dust shell and on the 15 $\mu$m
continuum peak. Four N4 IR sources show colors which  are consistent 
with reddened main sequence O stars, while the other three are IR objects.
 
From Figure \ref{Figure15} we can conclude the following:

- 1)  Star $\#1$ and $\#3$  are probably moderately reddened main sequence 
O stars with $\rm A_v \sim 3$. They are the main   ionizing stars 
responsible for the HII region (Heydari-Malayeri and Lecavelier des Etangs \cite{Heydari})
and labeled as A and B in Figure 1.
 
-2) Stars $\#5$ and $\#6$ are highly reddened main sequence O stars with 
$\rm A_v >10$. They do not have  optical counterparts.

-3) The IR objects labeled $\#4$, $\#5$, and $\#2$ have large IR excess and their colors 
are not that of reddened young main sequence stars. They lie in the border
of the molecular cloud seen in CO (Contursi \etal \cite{Contursi}) and at the edge of 
the cavity formed by the
ionizing stars. These three objects  have $J$ band brightness corresponding
to Class I and /or Herbig Ae/Be objects candidates (see Brandner \etal \cite{Bradner}).
Stars $\#7$ and  $\#4$ have the largest IR excess of the sample. They have IR characteristics
of   massive young Stellar Objects(YSO's) similar to N122 in 30Dor. 
Object $\#2$ is an extended object which shows nebulosity around and a second
red component.

\subsection{The nature of star $\#$ 4}

This is the brightest member of a small cluster and it is of particular interest because
it corresponds to
a bright point source emission in the hot dust pure continuum emission. Its
position in the NIR color-color and color-magnitude diagram is similar to that of
IR sources found in 30 Doradus, i.e. source W30, N122 and N126 (Rubio et al. 1998).
These 30 Doradus sources were resolved into multiple compact young massive systems from high resolution 
HST/NICMOS observations, with a dominant IR source (Walborn, et al 2002,
Walborn, Maiz-Apellaniz, Barba, 1999). The IR color of star \#4 can be either
interpreted as early O type star with  Av= 20 and with a IR K excess of about 4
magnitudes or a multiple object containing several   ZAMS O3  stars 
extinguished by 20 magnitudes. In fact, the star profile is wider than the point
spread function derived for the stars in the Ks image, which could be a hint of
the presence of a much redder component, thus favoring the hypothesis of a multiple system.

We can check the results on the nature of star $\#4$  obtained from NIR photometry,
with the comparison of our ISOCAM and NIR data with those 
published by  the ISOGAL team (Felli  \etal \cite{Felli}). The LW2 and LW3 broad band magnitudes 
obtained through PSF fitting are: m(LW2)= 6.1 and  
m(LW3)= 3.4. Then, the position of star $\#$4 on the [15]--[7]-[15] magnitude-color diagram published
by Felli \etal (\cite{Felli}) falls exactly in the region occupied by Young
Stellar Objects (YSOs).

A  third indication that  this objects contains  indeed an  YSO, is given by its 
CVF spectrum  shown in Figure \ref{Figure16}.
It shows a deep silicate absorption at $\sim 10$ $\mu$m which suggests this is a high embedded source.
This is also confirmed by the fact that star $\#4$ corresponds to the strongest peak of the 15 $\mu$m pure
continuum emission. In the spectrum are also visible  two weak absorption features 
at $\sim$ 13 $\mu$m  $\sim$15$ \mu$m.
We checked whether they are due to memory effects of the detector after a glitch. The result is that most
likely, the feature at 13 $\mu$m is fake but that at 15 $\mu$m  seems real.
If it is so, it could be associated with the solid CO$_2$ bending mode  at 15.2 
$\mu$m or to gas phase CO$_2$ at 15.0 $\mu$m. Both types of emission are detected in YSOs 
(van Dishoeck  \cite {vandishoeck}).  The fact that the overall spectrum shown 
in Figure  \ref{Figure16}
does not look like a typical YSO MIR spectrum (van Dishoeck  \cite {vandishoeck})  could be
due to the ISO spatial resolution; the ISOCAM beam  is in fact sufficiently large (2 pc at the LMC
distance) to include a considerable amount of  ISM surrounding the YSO, as
suggested  by the presence of AFEs. 
  To our knowledge this would be  the first time that the CO$_2$ absorption band is (marginally) 
detected in extragalactic YSOs, although SPITZER will certainly discover and study many more.
Since such  feature, together with the solid CO$_2$ stretch mode at 4.25 $\mu$m and H$_2$O at 3 $\mu$m 
(out of our wavelength range) has been observed at higher resolution with ISO-SWS in some Ultra Compact HII regions  
(Roelfsema \etal \cite{Roelfsema}) this would support  our hypothesis that star$\#$4 is an embedded YSO.

\begin{figure}
\includegraphics[width=0.5\textwidth,height=0.5\textwidth,origin=c,angle=-90]{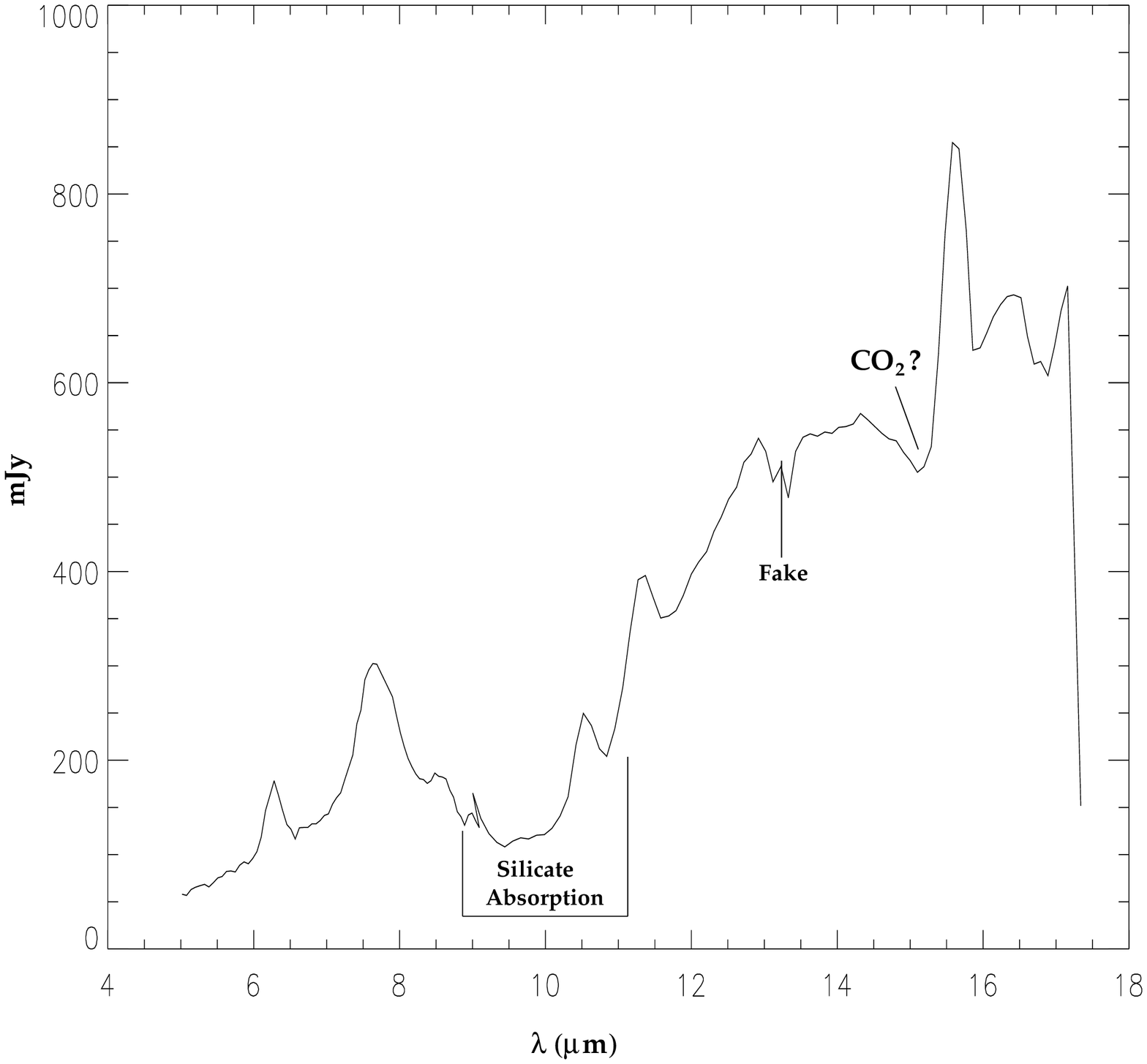}
\caption{ISOCAM CVF spectrum of star $\#$4 corresponding to the 15 $\mu$m pure continuum peak.}
\label{Figure16}
\end{figure}

%\newpage

\section{Conclusion}
 In this paper we have analyzed the ISOCAM CVF   and new $JHK_s$ photometry data of the
HII region complex N4 in the Large Magellanic Clouds.
The analysis has two principal aims: 1) to relate dust, gas and stellar content;
2) to look for effects of the lower than galactic metallicity on the dust features.
The main conclusions are the following.
\begin{itemize}
\item{The overall MIR spectral characteristics of N4 can be fully explained with the current dust models. The continuum + dust features + gas lines 
properties are those of a typical  galactic  HII  complex which can be interpreted as the combination of three primary components: 
 HII region core+ PDR+ parental molecular cloud.}
\item{Thanks to a fitting technique through the entire CVF data cube, we can produce maps in the single dust features, gas lines and pure 
continuum. We find that the HII region core is completely devoid of dust features. This dust cavity contains most of the ionized gas.
The pure continuum emission seems to have the same origin at all wavelengths, but it is probably arising from a population of grains different from the AFE carries.
Following the current dust models this component can be totally ascribed  to VSGs.}
\item{The comparison between the ISM emission and the stellar content of N4 derived from NIR photometric data, 
reveals that the two bluest stars are the main ionizing stars of N4A: they 
reside in the HII region core, {\it i.e.} in the center of the AFE cavity and where the ionized gas peaks. 
The pure continuum emission peaks on  a point source which 
shows strong evidences to be a  group of massive  young stars containing a deeply embedded YSOs.}
\item{A detailed comparison of the relative spatial distribution and intensities 
 of the dust features as function of the intensity of the radiation field,
suggests that the HII region core does not contain PAHs because they are destroyed. 
We show that this  
 mechanisms is  selective depending on the grains sizes, 
the smallest
grains/big--molecules, being destroyed first. This can lead to an overall  modification of 
the grains size distribution
 which can be significant  in  
relatively low metallicity dwarf galaxies, where these processes act on
large  portions of their ISM. We also argue that this mechanism is more important than 
any intrinsic dust modification due to the relatively low  metallicity of LMC.}
\item{The comparison between the   widths and the intensity ratios of the  dust features in N4 and in other galactic
environments shows that there are no    evident and significant signs of metallicity effects  on   AFE's carriers in N4}
\item{The analysis of the dust features intensities ratios and their comparison with model predictions, indicates that, if PAHs are the particles responsible for the
AFEs, they are fully ionized over the entire analyzed region.}
\item{The IR stellar content of the N4 shows several bright IR sources with
characteristics of massive early O type stars similar to those found in
30Doradus. IR spectroscopy of these sources would confirm their very young and
massive nature.}
\end{itemize}

\begin{acknowledgements} 
A.C. wishes to thank D. Lutz for his useful comments which significantly improved this paper.
\end{acknowledgements}

\end{document}